\documentstyle[preprint,eqsecnum,aps]{revtex}

\begin{document}

\draft
\preprint{HEP/123-qed}

\title{Metastable states and information propagation in
a 1D array of locally-coupled bistable cells}

\author{M.  P.  Anantram and Vwani P.  Roychowdhury\\}
\address{ University of California, Los Angeles, CA 90095-1594}

\date{\today}
\maketitle
\begin{abstract}
We study the effect of metastable states on the relaxation process (and
hence information propagation) in locally coupled and boundary-driven
structures. We first give a general argument to show that metastable
states are inevitable even in the simplest of structures, a wire. At
finite temperatures, the relaxation mechanism is a thermally assisted
random walk. The time required to reach the ground state and its
life time are determined by the coupling parameters. These time scales
are studied in a model based on an array of quantum dots. 
\end{abstract}

\pacs{  }

A number of novel proposals have recently been advanced, articulating
primarily, visions of future computing systems~\cite{Bakshi91,PhysComp94,Bandyopadhyay94,Korner93,Llyod93,Cirac95,Glanz95} 
using nanoelectronic structures~\cite{Likharev92,Kastner96,Andres96}, 
in which computation is fundamentally related to the underlying physics
of the devices.
In this paper, we consider one such class of models comprising of an
array of locally-coupled and edge-driven cells, where computation is
realized by relaxation of the physical system to its ground
state~\cite{Bakshi91,PhysComp94,Bandyopadhyay94}. These models are
semiclassical and global phase coherence is not maintained as the
system relaxes to its ground state by dissipative processes.

The computing architecture comprises of locally-coupled arrays of a
basic unit that exhibits bistability and the bistable states are used
to represent the binary values 0 and 1 (Fig. 1(a)).
The basic units interact with their nearest neighbors to form larger
devices such as wires, logic gates and cellular automata (Fig.
1(b))~\cite{PhysComp94,Bandyopadhyay94}. The units on 
the edges form the input and output ports (Fig. 1(b)),
and the interior units of the device are not externally
accessed~\cite{PhysComp94,Bandyopadhyay94}. Both information and 
energy are
provided to these boundary ports as input data. {\it Central} to the
operation of the computing system is the assumption that the system
then relaxes by dissipation to the ground state that depends only on
the configuration of the input cells, which are held fixed. The ground
state configuration represents the result of the computation and the
output is read from the cells marked `output' (Fig. 1).
For example, consider a wire which
consists of a linear array of basic units as shown in Fig. 2(a).
The ground state is two fold degenerate; all
cells have a bit value of either 0 or 1 (Fig. 2(b)).
Computation is based on the thesis that the system
always relaxes to the ground state determined by the bit value of the
input cell, thus transmitting information from the input to the output
of the wire.

The proposed models, which involve a Hubbard-type Hamiltonian clearly
demonstrate that the ground state of suitably designed structures
correspond to computationally useful 
operations~\cite{PhysComp94,Bandyopadhyay94}. While this is an 
important first step, it is essential to study the dynamical evolution 
of the computational trajectory because it is a priori not clear that 
the system will in fact relax to the ground state ever, or, in a time
efficient manner. Our specific concern is metastable states, which may
hinder relaxation to the ground state as hypothesized in Refs.
\onlinecite{Bandyopadhyay96} and \onlinecite{Landauer_meta}.

In the first part of the paper, we will give a simple insightful
argument as to why metastable states are inevitable in a wire. As a
 consequence of the metastable states, information propagation is
not feasible at extremely low temperatures. The computing system can
however escape from the metastable states and reach the ground state
at non zero temperatures via a thermally assisted random walk; thus 
propagating information from the input to the output end of the wire. 
A discussion of this in the context of a model comprises the second 
part of the paper.

Consider the initial state of a wire with bit value 0 in the input
cell and bit value 1 in the remaining cells (Fig. 2(a)). 
The intersection of the left and right aligned cells is referred
to as a {\it kink}. If the system relaxes to the ground state as time
evolves, it is expected that the kink will propagate towards the right,
into the bulk of the wire. The state where the kink is $m$ units from
the input end is denoted by $\phi_m$ and its energy is
denoted by $E(m)$ (Fig. 3(a)). The computational
trajectory from $\phi_m$ to $\phi_{m+1}$ may in principle involve many
intermediate states but for simplicity we will assume that there is
only a single intermediate state which is denoted by $\phi_{m}^\prime$
and its energy is denoted by $E^\prime(m)$. In a long wire, states
$\phi_m$ and $\phi_{m+1}$ have almost the same energy because their
left and right environments are nearly the same. As a result, the total
energy versus state of the system should vary periodically in the bulk
as shown in Fig. 3(b). $\phi_m$ is a metastable state
because the $\phi_m \rightarrow \phi_m^\prime$ and $\phi_m \rightarrow
\phi_{m+1}^\prime$ transitions  require surmounting of energy barriers
in the bulk ($\Delta E$).  Note that in principle the energies of the
primed and unprimed states could be interchanged, in which case the
primed states will be the metastable states.

Varying the inter and intra cell interactions yields different values
of the $\Delta E$. However, even for parameters where $\Delta E$ is
close to zero, we argue that metastable states should exist at least
either
near the left or right edges. The cells close to the edges
have a non symmetric environment to its left and right. As a result,
$\Delta E$ will be non zero near the edges. If $E^\prime(1) > E(1)$,
then trivially the initial state is a metastable state. If not, then
by symmetry of the structure $E(N-1) > E^\prime(N-2)$ and this results
in a metastable state ($N$ is the total number of
cells in the wire). It should also be noted that in a structure
where $\Delta E$ is designed to be zero, small undesired fluctuations
in the inter and intra cell interactions cause $E(m) \neq E(m+1)$,
leading to metastable states~\cite{Landauer_meta}.

We have shown that metastable states exist at zero temperature. At
finite temperatures, the kink propagates to the right via a thermally
assisted random walk, where the various states, accessed (Fig. 3)
from the input end to the output form the lattice
points. Note that the thermally assisted random walk is the only
mechanism to overcome any $\Delta E$ since there are no fields driving
the computation. The probability to hop from the lattice point
representing state $\phi_m$ to the lattice points representing
$\phi_m^\prime$ and $\phi_{m-1}^\prime$ are given by
${\Gamma_{\phi_m \rightarrow \phi_m^\prime}}/({\Gamma_{\phi_m
\rightarrow \phi_m^\prime} + \Gamma_{\phi_m \rightarrow
\phi_{m-1}^\prime}})$ and ${\Gamma_{\phi_m \rightarrow
\phi_{m-1}^\prime}}/({\Gamma_{\phi_m \rightarrow \phi_m^\prime} +
\Gamma_{\phi_m \rightarrow \phi_{m-1}^\prime}})$ respectively, where
$\Gamma_{a \rightarrow b}$ is the transition rate to go from state $a$
to state $b$. A similar expression applies for the transition
probabilities from $\phi_m^\prime$ to $\phi_{m}$ and $\phi_{m+1}$. For
a uniform and long wire, these transition probabilities are equal to
one half, when $\phi_m$ represents a state where the signal (kink) has
traveled into the interior of the wire. This is because the energy
differences $E^\prime(m) - E(m)$ and $E^\prime(m-1) - E(m)$ are nearly
the same.  The random walk is a finite one with an absorbing boundary
on the right end (the walk stops when the output is reached) and a
reflecting boundary on the left end (signifies that the input cell has
a fixed state). The average time taken for the signal to propagate from
the initial state to the ground state ($T_{tot}$) is the quantity of
interest. We are not aware of analytical techniques to calculate
$T_{tot}$. However, to understand the underlying  physics,  it is
useful to consider the time required for the kink to travel $n$ cells
in the bulk of a wire (i.e. edge effects are not included). This time
follows directly from the discussion of random walk~\cite{Reif65} and
is,
\begin{eqnarray}
\mbox{Time required to travel $n$ cells} \sim (2n)^2\;
\frac{\tau_1+\tau_2}{2}
                                \mbox{ , } \label{eq:diffusion}
\end{eqnarray}
where, $\tau_1$ and $\tau_2$ are the life times of the metastable 
states and the states in between two consecutive metastable states
(Fig. 3). The factor $2n$ represents the fact that 
there are two states involved in the propagation of the signal across 
each of the $n$ cells. At low temperatures ($kT < \Delta E$), the time
in Eq. (\ref{eq:diffusion}) is primarily determined by $\tau_1 \sim
\tau_0 exp({\Delta E}/{kT})$, where $\tau_0$ and $\Delta E$, the 
energy barrier of the metastable states encountered in the bulk (Fig. 
3), depend on the particular model. From the 
exponential dependence of $\tau_1$,
it is clear that $T_{tot}$ increases exponentially with decrease in 
$kT$.  $T_{tot}$ can be made smaller by raising the temperature. The
temperature, however, cannot be increased indefinitely because for the
computation to be useful, the system must remain in the ground state
for a long enough time so that we know for certain that the ground
state has been reached and that the system has not escaped from it.
The life time in the ground state varies with temperature as $\tau_g
exp(\Delta E_g/kT)$, where $\tau_g$ and $\Delta E_g$
(the energy barrier separating the ground state and the next excited
state that can be reached) are constants.
To realize a computation, it is important that the
temperature is large enough to shake the system out of a metastable
state but not so large as to excite the system out of the ground state
in a short time period. Hence, it is necessary that the energy
difference between the ground state and the excited states (which are
reached by a single electron tunneling event from the ground state) is
many times the thermal energy $kT$ and the energy barrier due to
metastable states is comparable to or smaller than $kT$. The relative
magnitudes of $\Delta E$, $\Delta E_g$ and other energy barriers close
to the edges are determined by the device parameters. It is
essential to determine if there exists a region in the parameter space
of the device dimensions and temperature, where the system relaxes to
the ground  state quickly and also remains there for a sufficiently
long time period to be computationally useful. We study this by
considering the time evolution of a specific model.

The model considered is one similar to that discussed in the literature
before~\cite{Lent94}. The basic unit consists of four identical
metallic type quantum dots containing a net charge of two electrons. A
wire is comprised of a linear array of these units
(Fig. 4). The tunnel resistance between the dots is much
larger than ${h}/{e^2}$. In this limit, each dot contains an
integer number of electrons. The intra and inter cell interactions are
modeled by capacitances $C$ and $C^\prime$ respectively. $C_0$ is the
capacitance between a dot and the ground. The values of these
capacitances do not vary from cell to cell. The tunnel resistances
between the dots along the sides of a single cell are represented by 
$R$ and the tunnel resistances between
dots along the diagonals of a cell and between dots of different cells
are infinite. Though imaging of charge to the outside world is
unavoidable, models for ground state computing have neglected this
feature all together. The total energy of a charged system without
imaging to the outside world (i.e., $C_0=0$) is infinite and so we have
to assume at least a small value for $C_0$ in the calculations. We have
chosen $C_0 = 0.001 C$ throughout this paper.

In the lowest energy state, the two excess electrons in
an isolated cell are aligned either along the left or right diagonals
and these two states represent the binary values 0 and 1. The ground
state of the wire is two fold degenerate as shown in Fig. 2(b).

The time evolution is modeled by the orthodox theory of single
electron tunneling~\cite{Averin91}, where the transition
rate for a single electron tunneling event from
dot $i$ to dot $j$ is,
\begin{eqnarray}
\Gamma_{ij} = \frac{\Delta E_{ij}}{q^2 R} \frac{1}
{exp(\Delta E_{ij}/kT) - 1} \mbox{,}
\label{eq:Trans_Probab}
\end{eqnarray}
where $R$ is the tunnel resistance between dots $i$ and $j$.
$\Delta E_{ij} = E_a - E_b$, where $E_a$ and $E_b$ are total energies
of the system after and before the tunneling event.
Using the standard monte carlo method as applied to single electron
tunneling~\cite{Likharev1}, we compute the time taken to reach the
ground state ($T_{tot}$) as a function of the temperature for wires of
various lengths, with the initial states as shown in Fig.
2(a). The main results of this simulation are summarized
in Fig.  5(a). As the temperature tends to zero, $T_{tot}$
tends to infinity and the ground state is never reached (this is not
strictly true if {\it higher order} quantum mechanical co-tunneling
processes are included).  At small temperatures, where $kT$ is much
smaller than all barrier heights encountered in the computational
trajectory, $T_{tot}$ decreases exponentially with increase in
temperature. This can be understood by noting that the various
tunneling probabilities depend exponentially on ${\Delta E}/{kT}$,
where $\Delta E$ is the barrier that an electron should surmount to
overcome a metastable state (Eq. (\ref{eq:Trans_Probab})). At
temperatures comparable to or larger than $\Delta E$, $T_{tot}$
decreases inversely with temperature. This can again be understood from
Eq. (\ref{eq:Trans_Probab}) because for $kT > \Delta E$,  $\Gamma_{ij}
\; \propto \; T^{-1}$. The temperature however cannot be made very
large because the life time of the system in the ground state
decreases with increase in temperature and this is undesirable for
computing. The largest temperatures chosen in Fig. 5(a) is
$0.075 {e^2}/{C}$, which is larger than $\Delta E_g$  (Fig. 3). 
Even at this temperature, which is not suited for
ground state computing, the time taken to reach to ground state is too
large; it takes $10^5 RC$ for the system to reach the ground state of
a wire with only sixty cells. 
A value of ${C^\prime}/{C}=1.3$ was chosen for these simulation. We 
have performed simulations for other parameters and find that the
results are not significantly different. 
Though Eq. (\ref{eq:diffusion}) was not intended to calculate $T_{tot}$,
we remark that substituting $n=60$ and the appropriate values for $kT$
and $\Delta E$ in Eq. (\ref{eq:diffusion}) gives a time which agrees 
with $T_{tot}$ to about an order of magnitude.

To see that there is only a narrow region of capacitance parameters
and temperature where the system relaxes to the ground state and
remains there for long times, we plot  $\Delta E$ and $\Delta E_g$
versus ${C^\prime}/{C}$ in Fig. 6. Here, for
${C^\prime}/{C} \ge 2.1$, $|\Delta E| \ge |\Delta E_g|$. This leads
to a shorter life time in the ground state than in the metastable
states and is therefore undesirable for computing. For
${C^\prime}/{C} \le 1$, the magnitude of $|\Delta E_g|- |\Delta E|$
decreases as ${C^\prime}/{C}$ becomes smaller. As a result, the 
temperature of operation has to be made correspondingly smaller to 
ensure that the life time in the ground state is significantly larger
than the life time in the metastable states. However, $T_{tot}$ 
increases exponentially with decrease in temperature and so very small 
values of ${C^\prime}/{C}$ are also undesirable. Fig. 6 is
a plot of only two types of metastable states, the ones in the bulk of 
the wire and $\Delta E_g$. Other metastable states close to the edges 
of wire play an important role too, especially for small values of 
$|\Delta E|$.

Finally, we address the issue of how the system evolves to the ground
state: A wire with $N$ cells has $8(N-1)$ possible transitions which
compete to determine the state of the wire in the next time step. In
the discussion surrounding Fig. 3, we assumed that
the cells flip from the left  (input) to the right end in a specific
manner such that a single kink performs a thermally assisted random
walk. We find from our monte carlo simulations that this picture is
valid: having more than one kink is energetically expensive and each
cell flips from a left to a right aligned one (or vice versa) via the
path shown in Fig. 5(b).

We have also performed simulations on a fanout
gate~\cite{PhysComp94,Bandyopadhyay94}, which comprises of a single 
input, which is transmitted to two output ports (Fig. 7).
The metastable states that have to be overcome here are far worse
and an intuitive reason is as follows. The polarization of Cell 2 can
flip as in the case of a wire. For the signal to propagate further,
it is essential for Cell 3 to flip its polarization. Cell 3
has two neighbors (4 and 7) which have the same polarization and one
neighbor (Cell 2) with the opposite polarization. It is however
energetically unfavorable for Cell 3 to flip because the new state
would have two kinks, thus resulting in a far worse metastable state
than in the case of the wire. In summary, we have shown that metastable
states always exist in locally-coupled edge-driven computing systems, 
thus preventing information propagation at very low temperatures. At 
finite temperatures, relaxation to the ground state and hence 
propagation of information along the wire takes place by the 
inefficient process of a thermally assisted random walk.

\noindent
{\it Note:} We would also like to bring to notice a recent
proposal~\cite{Lent96} for computing which uses bistable cells evolving
under the presence of a spatially and adiabatically time varying field.
The physics of such systems are different from that discussed here.

This work was supported in part by NSF Grants No. ECS-9308814 and No.
ECS-9523423.




\pagebreak

FIGURE CAPTIONS:

Fig. 1:
(a) A single bistable cell: the lines along the left and right
diagonals represent the binary values 0 and 1.
(b) schematic of a corresponding computing system.

\vspace{0.1in}

Fig. 2: 
(a) The initial state and (b) the two degenerate ground states of the wire.

\vspace{0.1in}

Fig. 3: 
(a) State $\phi_m$, where the signal has transmitted $m$ cells deep
into the wire.
(b) A qualitative plot of the energy versus the states accessed as
the signal propagates away from the edges of the wire.

Fig. 4:
(a) Electrons occupying diagonally opposite dots represent binary
values of 0 and 1.
(b) A wire constructed from the basic units in (a).

\vspace{0.1in}

Fig. 5:
(a) Time taken to reach the ground state versus temperature for wires 
of different lengths ($C^\prime/C =1.3$)
(b) Sequence of steps by which the polarization of a cell flips.

\vspace{0.1in}

Fig. 6: 
A plot of $|\Delta E|$ (solid line) and $|\Delta E_g|$ (dashed line)
discussed in the context of Fig. 4 versus ${C^\prime}/{C}$.

\vspace{0.1in}

Fig. 7: 
Fanout gate

\vspace{-0.2in}

\begin{thebibliography}{100}


\bibitem{Bakshi91}
P. Bakshi, D.~A. Broido, and K. Kempa, J. Appl. Phys. {\bf 70},  5150
(1991).

\bibitem{PhysComp94}
C. Lent, P.~D. Tougaw, W. Porod, PhysComp '94, p. 5-13, (IEEE Press, 
1994); 
C. Lent, P.~D. Tougaw, W. Porod, and G.~H. Bernstein, Nanotech.
{\bf 4},  49 (1993).

\bibitem{Bandyopadhyay94}
S. Bandyopadhyay, B. Das, and A.~E. Miller, Nanotechnology {\bf 5},
113 (1994).

\bibitem{Korner93}
H. Korner and G. Mahler, Phys. Rev. B {\bf 48},  2335  (1993).

\bibitem{Llyod93}
S. Llyod, Science {\bf 261},  1569  (1993).

\bibitem{Cirac95}
J.~I. Cirac and P. Zoller, Phys. Rev. Lett. {\bf 74},  4091  (1995).

\bibitem{Glanz95}
J. Glanz, Science {\bf 269},  1363  (1995).

\bibitem{Likharev92}
K. K. Likharev and T. Claeson, Scientific American {\bf 266},
50 (1992) and references there in.

\bibitem{Kastner96}
M. A. Kastner, Comm. in Cond. Matt. Phys. {\bf 17}, 349  (1996)
and references there in.

\bibitem{Andres96}
R. P. Andres, J. D. Bielefeld, J. I. Henderson, D. B.
Janes, V. R. Kolagunta, C. P. Kubiak, W. J. Mahoney
and R. G. Osifchin, Science {\bf 273}, 1690 (1996).

\bibitem{Bandyopadhyay96}
S. Bandyopadhyay, Preprint.

\bibitem{Landauer_meta}
R. Landauer, Phil. Trans. R. Soc. Lond. A {\bf 353}, 367 (1995).

\bibitem{Reif65}
F.~Reif, {\it Fundamentals of Statistical and Thermal Physics}
(McGraw-Hill, U.S.A, 1965).

\bibitem{Lent94}
C. Lent and P.~D. Tougaw, J. of App. Phys. {\bf 75},  4077  (1994).

\bibitem{Averin91}
D.~V. Averin and K.~K. Likharev,  in {\em Mesoscopic Phenomena in
Solids}, edited by B.~L. Altshuler, P.~A. Lee, and R.~A. Webb (Elsevier
Science Publishers, U. S. A, 1991).

\bibitem{Likharev1}
N.~S. Bakhalov, G.~S. Kazacha, K.~K. Likharev, and S.~I. Serdyokova,
Sov. Phys.  JETP {\bf 68},  581  (1989).

\bibitem{Lent96}
C. Lent, P.~D. Tougaw, W. Porod, PhysComp '96, p.186 (Eds.
T. Toffoli, Michael Biafore and Joao Leao, New England Complex Systems 
Institure, 1996)
\end{thebibliography}
\end{document}